\documentclass[journal,twoside,web]{ieeecolor}
\usepackage{lcsys}
\usepackage{cite}
\usepackage{amsmath,amssymb,amsfonts}
\usepackage{algorithmic}
\usepackage{graphicx}
\usepackage{textcomp}
\usepackage{xcolor}

 \usepackage[italian]{babel}
 \usepackage{pstricks}
 \usepackage{verbatim}
 \usepackage{graphicx}
 \usepackage{psfrag}
 \usepackage{fancybox}
 \usepackage{amsmath}
 \usepackage{amsfonts}
 \usepackage{amssymb}
 \usepackage{listings}
 \usepackage{enumerate}
 \usepackage{url}

\input{pst-plot}
\input{pst-node}
\input{pst-coil}
 \psset{unit=1.0\unitlength}
 \SpecialCoor

\newcommand{\schemaPOG}[3]{
 \setlength{\unitlength}{#2}
 \psset{unit=1.0\unitlength}
 \SpecialCoor
 \begin{pspicture}#1
 \put(0,0){$#3$}
 \end{pspicture}
}

\newcommand{\punto}{\pscircle*{0.16}}

\newcommand{\tratteggio}{\psline[linestyle=dashed,linewidth=0.2pt](0,-0.75)(0,10.75)}

\newcommand{\piu}{\begin{pspicture}(0,0)
 \put(0,0){\pscircle[linewidth=0.4pt,fillstyle=none]{0.5}}
 \psline(0.5; 45)(0.5;225)
 \psline(0.5;315)(0.5;135)
\end{pspicture}}

\newcommand{\dst}{\begin{pspicture}(0,0)
 \put(0.293,0){\pscircle*{0.207}}
\end{pspicture}}

\newcommand{\snt}{\begin{pspicture}(0,0)
 \put(-0.293,0){\pscircle*{0.207}}
\end{pspicture}}

\newcommand{\sumdstsnt}[1]{\begin{pspicture}(0,1)
   \put(4,0){\piu}
   \put(4,0){#1}
   \put(0,0){\psline{->}(3.5,0)}
   \put(8,0){\psline{->}(-3.5,0)}
\end{pspicture}}

\newcommand{\dotsntdst}{\begin{pspicture}(0,1)
   \put(4,0){\punto}
   \put(4,0){\psline{->}(-4,0)}
   \put(4,0){\psline{->}(4,0)}
\end{pspicture}}

\newcommand{\midgiu}[2]{\begin{pspicture}(0,1)
   \put(-2,3){\psframe(4,4)\rput(2,2){$#1$}}
   \put(0,3){\psline{->}(0,-3)}
   \put(0,9.5){\psline{->}(0,-2.5)}
   \put(0,-0.5){\makebox(0,0)[t]{$#2$}}
\end{pspicture}}

\newcommand{\midsu}[2]{\begin{pspicture}(0,1)
   \put(-2,3){\psframe(4,4)\rput(2,2){$#1$}}
   \put(0,7){\psline{->}(0,3)}
   \put(0,0.5){\psline{->}(0,2.5)}
   \put(0,10.5){\makebox(0,0)[b]{$#2$}}
\end{pspicture}}

\newcommand{\blosubase}[2]{\begin{pspicture}(8,10)
   \put(0, 0){\tratteggio}
   \put(0,10){\dotsntdst}
   \put(4, 0){#1}
   \put(0, 0){\sumdstsnt{#2}}
\end{pspicture}}

\newcommand{\blogiubase}[2]{\begin{pspicture}(8,10)
   \put(0, 0){\tratteggio}
   \put(0,10){\sumdstsnt{#2}}
   \put(4, 0){#1}
   \put(0, 0){\dotsntdst}
\end{pspicture}}

\newcommand{\bloin}[2]{\begin{pspicture}(0,0)
 \put(-0.5,10){\makebox(0,0)[r]{$#1$}}
 \put(-0.5, 0){\makebox(0,0)[r]{$#2$}}
 \put(0, 0){\punto}
 \put(0,10){\punto}
 \end{pspicture}}

\newcommand{\bloout}[2]{\begin{pspicture}(0,10)
   \put(0.5,10){\makebox(0,0)[l]{$#1$}}
   \put(0.5, 0){\makebox(0,0)[l]{$#2$}}
   \put(0, 0){\punto}
   \put(0,10){\punto}
   \put(0,0){\tratteggio}
\end{pspicture}}

\newcommand{\blosu}[3]{\blosubase{\midsu{#1}{#2}}{#3}}

\newcommand{\blogiu}[3]{\blogiubase{\midgiu{#1}{#2}}{#3}}

\newcommand{\kxdst}[4]{\begin{pspicture}(0,1)
   \put(0,0){\psline{->}(1,0)}
   \psset{unit=0.5\unitlength}
   \rput[lb](2,-#1){\psset{unit=2.0\unitlength}
   \framebox(#1,#1){$#2$}}
   \psset{unit=2.0\unitlength}
   \put(#1,0){
   \put(1,0){\psline{->}(1,0)}
   \put(2.25, 0.5){\makebox(0,0)[bl]{$#3$}}
   \put(2.25,-0.5){\makebox(0,0)[tl]{$#4$}}}
\end{pspicture}}

\newcommand{\kxsnt}[4]{\begin{pspicture}(0,1)
   \put(1,0){\psline{->}(-1,0)}
   \put(-0.25, 0.5){\makebox(0,0)[br]{$#3$}}
   \put(-0.25,-0.5){\makebox(0,0)[tr]{$#4$}}
   \put(#1,0){
   \put(2,0){\psline{->}(-1,0)}
   \psset{unit=0.5\unitlength}
   \rput[rb](2,-#1){\psset{unit=2.0\unitlength}
   \framebox(#1,#1){$#2$}}
   \psset{unit=2.0\unitlength}}
\end{pspicture}}

\newcommand{\blokxbase}[3]{\begin{pspicture}(#1,10)
   \put(0, 0){\tratteggio}
   \put(0, 0){#2}
   \put(0,10){#3}
\end{pspicture}
\begin{pspicture}(2,10)\put(0, 0){}\end{pspicture}
}
\newcommand{\kduedst}[3]{\kxdst{2}{#1}{#2}{#3}}
\newcommand{\kduesnt}[3]{\kxsnt{2}{#1}{#2}{#3}}

\newcommand{\blokao}[4]{\blokxbase{2}{\kduedst{#2}{}{#4}}{\kduesnt{#1}{#3}{}}}

\newcommand{\flowright}[1]{\begin{pspicture}(0,12)
   \put(0,-1){\makebox(0,0)[t]{$\Rightarrow$}}
\end{pspicture}}

\newcommand{\flowleft}[1]{\begin{pspicture}(0,12)
   \put(0,-1){\makebox(0,0)[t]{$\leftarrow$}}
\end{pspicture}}

\newcommand{\tras}{^{\mbox{\tiny T}}}

\newcommand{\ds}{\displaystyle}

 \newcommand{\mat}[2]{\left[\begin{array}{#1} #2 \end{array}\right]}

   \def\h{{\bf h}}
   
   \def\p{{\bf p}}

  \def\cS{{\cal S}}

 \newtheorem{Algo}{\bf Algorithm}
 \newtheorem{Theo}{\bf Theorem}
 
 \newtheorem{Prop}{\bf Property}
 \newtheorem{Def}{\bf Definition}

 \newcommand{\eps}{.}

\def\BibTeX{{\rm B\kern-.05em{\sc i\kern-.025em b}\kern-.08em
    T\kern-.1667em\lower.7ex\hbox{E}\kern-.125emX}}
\begin{document}
\title{Residues in Partial Fraction Decomposition Applied to Pole Sensitivity Analysis\\ and Root Locus Construction}
\author{
Davide Tebaldi \IEEEmembership{Member, IEEE}
and
Roberto Zanasi \\
This work has been submitted to the IEEE for possible publication. Copyright may be transferred without notice, after which this version may no longer be accessible. \vspace{-6.8mm}
\thanks{The work was partly supported by the University of Modena and Reggio Emilia
through the action FARD (Finanziamento Ateneo Ricerca Dipartimentale) 2023/2024, and funded under the National Recovery and Resilience Plan (NRRP), Mission 04 Component 2 Investment 1.5 - NextGenerationEU, Call for tender n. 3277 dated 30/12/2021
Award Number:  0001052 dated 23/06/2022.}
\thanks{Davide Tebaldi and Roberto Zanasi are
with the Department of Engineering ``Enzo Ferrari'', University of
Modena and Reggio Emilia, Modena, Via Pietro Vivarelli 10 - int. 1 -
41125 Modena, Italy. E-mails: davide.tebaldi@unimore.it, roberto.zanasi@unimore.it.}
}

\maketitle



\begin{abstract}
The applications of the partial fraction decomposition in control and systems engineering are several. In this letter, we propose a new interpretation of residues in the partial fraction decomposition, which is employed for the following purposes: to address the pole sensitivity problem, namely to study the speed of variation of the system poles
when the control parameter changes and when the system is subject to parameters variations,
as well as to propose a new algorithm for the construction of the root locus. The new algorithm is proven to be more efficient in terms of execution time than the dedicated MATLAB function, while providing the same output results.
\end{abstract}


\section{Introduction}
\label{sec:introduction}

\IEEEPARstart{T}{he} partial fraction decomposition technique is a powerful calculus tool
in the field of control and systems engineering, which allows to break down a rational function into a sum of simpler fractions. This method is particularly useful to integrate rational functions and to solve differential equations, and
finds many different applications. As an example, it is employed in \cite{Nuovo_1} to  decompose the transfer function of infinite impulse response filters into the sum of fractions, which is ultimately used to perform the optimal design of the filters. In \cite{Nuovo_2}, the partial fraction decomposition concept is used as a basis for an adaptive Antoulas-Anderson algorithm, which allows to derive an approximation of the transfer function starting from the data acquisitions of the frequency response.
Other uses of the partial fraction expansion include algorithms to compute the storage function in lossless systems \cite{Nuovo_3}, and the calculation of the Padè approximation of transfer functions via a Lanczos process \cite{Nuovo_4}.

Many different methods have been developed over the years for the computation of the residues~\cite{PFE_1}, offering different pros and cons with respect to each other. Out of these methods, the so-called cover-up method is the one which is typically employed in control theory~\cite{LTI_Syst_1}.

\begin{figure}[t!]
\centering
 \setlength{\unitlength}{2.0mm}
 \psset{unit=1.0\unitlength}
 \begin{picture}(44,9.5)(2,0)
 \put(3,5){\vector(1,0){2}}
 \put(5,3){\framebox(4,4){$\ds K $}}
 \put(9,5){\vector(1,0){2}}
 \put(11.5,5){\piu}
 \put(11.5,5){\snt}
 \put(11.5,8){\punto}
 \put(11.5,8.0){\vector(0,-1){2.5}}
 \put(11.5,8.5){\makebox(0,0)[b]{\small$r(t)$}}
 \put(12,5){\vector(1,0){2}}
 \put(14,3){\framebox(7,4){$\ds G(s)$}}
 \put(21,5){\vector(1,0){4.5}}
 \put(23.5,5){\line(0,-1){4}}
 \put(23.5,1){\line(-1,0){20.5}}
 \put(3,1){\line(0,1){4}}
 \put(24,5.5){\makebox(0,0)[b]{\small$y(t)$}}
 \put(26.8,3){$\Rightarrow$}
 \put(30.5,3.5){\vector(1,0){2}}
 \put(31.5,4){\makebox(0,0)[br]{\small$r(t)$}}
 \put(32.5,1.5){\framebox(7,4){$\ds G_0(s)$}}
 \put(40.5,4){\makebox(0,0)[lb]{\small$y(t)$}}
 \put(39.5,3.5){\vector(1,0){2}}
 \end{picture}
 \vspace{-6mm}
\caption{Linearly controlled feedback system.}
\label{Linearly_controlled_system}
 \vspace{-3mm}
\end{figure}
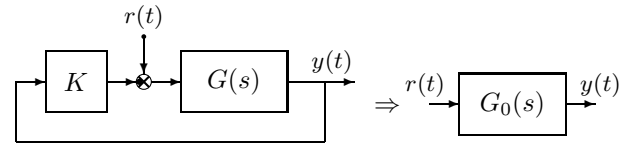

However, the actual meaning of residues has not been fully investigated in the literature yet. In this letter, we investigate the meaning of residues in detail, finding a remarkable result: given an open loop system described by the transfer function $G(s)$, from which a linearly controlled feedback system can be constructed to obtain the closed-loop transfer function $G_0(s)$ as illustrated in Fig.~\ref{Linearly_controlled_system}, the residues of the partial fraction decomposition of $G_0(s)$ taken with opposite signs are exactly the speed of variation of the poles of the closed-loop system as a function of the control parameter $K$.
This result introduces different potential applications, out of which we propose three in this letter: 1) The pole sensitivity analysis of the feedback system to variations of the control parameter $K$, that is the study of how fast the system poles move in the complex plane as a function of the control parameter $K$, which represents an extension of the root locus method~\cite{Nuovo_5,Nuovo_11}. The root locus method is known to be widely employed both in classical automatic control as well as in more advanced control theory applications, as in \cite{Nuovo_9} and \cite{Nuovo_10} where it is applied to multi-agent systems. Its applications also extend to other engineering fields, such as the use of the root locus for the analysis and control design of power systems~\cite{Nuovo_6}, and the optimal design of delta–sigma modulators~\cite{Nuovo_7}, for instance.
However, the branches of the root locus simply give information about the location of the poles as a function of the control parameter $K$, without giving any specific information about the speed of variation of the poles themselves. In order to derive the latter piece of information, the derivative of the root locus branches should be computed, which can lead to inaccuracy since the derivative is computed numerically. The analysis proposed in this letter allows to determine a very simple analytical and exact way of computing the speed of variation of the poles in the complex plane, which coincide with the residues of the partial fraction decomposition of the closed-loop transfer function taken with opposite signs.
2) Based on the performed analysis, we propose a new algorithm for the derivation of the root locus. The derivation of the root locus is normally performed using the tools available from computer aided control system design software, such as the {\tt rlocus} MATLAB function for example~\cite{Nuovo_8}.
The algorithm proposed in this letter is proven to work more efficiently than the {\tt rlocus} MATLAB function in terms of execution time, while providing the same output results.
3) The pole sensitivity analysis of the system when subject to parameters variations. The three previous applications have then been applied to different dynamical system case studies.

The remainder of this letter is structured as follows. The definition of pole sensitivity is given in Sec.~\ref{Poles_Sensitivity_of_a_Dynamic_System_sect}, and investigated in Sec.~\ref{Poles_sensitivity_of_a_linearly_controlled_feedback_system_sect} for a feedback system. The new algorithm for the root locus calculation is proposed in Sec.~\ref{Algorithm_for_Root_locus_calculation_sect}, while Sec.~\ref{Poles_sensitivity_of_a_dynamic_system_to_its_own_parameters_sect} addresses the pole sensitivity analysis of a system subject to parameters variations. The conclusions are given in Sec.~\ref{Conclusion_sect}.

\section{Poles Sensitivity of a Dynamical System}\label{Poles_Sensitivity_of_a_Dynamic_System_sect}

\begin{Def}\label{definition_sensitivity}
We define the term {\it pole sensitivity of a dynamical system} as the study of how fast
the poles of a dynamical system move in the complex plane as the system's parameters change.
\end{Def}

In this letter, the following types of {\it pole sensitivity} will be considered:
1) pole sensitivity of a linearly controlled feedback system.
2) pole sensitivity of a dynamical system to its own parameters.

\section{Pole sensitivity of a linearly controlled feedback system}\label{Poles_sensitivity_of_a_linearly_controlled_feedback_system_sect}

 Reference is made to the feedback system of Fig.~\ref{Linearly_controlled_system}, where $K$ is a proportional controller and $G(s)$ is the rational transfer function of the considered dynamical system:
\begin{equation}
\label{Gs_function}
G(s)=\frac{N(s)}{D(s)}=\frac{N(s)}
{(s-p_1)\,(s-p_2)\,\ldots\,(s-p_n)},
\end{equation}
where $p_j$ are the poles of function $G(s)$, for $j\in\{1,\,2,\,\ldots,\,n\}$.  In the case of simple poles, the function $G(s)$ can always be decomposed into partial fractions as follows:
\[G(s)= \sum_{j=1}^{n} \frac{\bar{k}_{j}}{s-p_{j}},\]
where $\bar{k}_{j}$ are the residues of function $G(s)$ at the poles $s=p_{j}$:
\begin{equation}
\label{Residues}
\bar{k}_{j}=\frac{N(p_j)}{\prod_{h=1,h\not=j}^{n}(p_{j}-p_{h})}.
\end{equation}
Note: the residues $\bar{k}_{j}$ are real numbers if pole $p_j$ is real, and are complex numbers if pole $p_j$ is complex. It is a well known result~\cite{LTI_Syst_1} that the poles of the feedback system  in Fig.~\ref{Linearly_controlled_system} are the poles of the following closed-loop transfer function $G_0(s)$:
\begin{equation}
\label{G0s}
G_0(s)
=\frac{Y(s)}{R(s)}
=\frac{N(s)}{D(s)+K\,N(s)}
=\frac{N(s)}{\Delta(s,K)},
\end{equation}
namely the roots of the following characteristic equation:
\begin{equation}
\label{equazione_caratteristica}
\Delta(s,K)=D(s) + K N (s) =0,
\end{equation}
where $N(s)$ and $D(s)$ are the numerator and the denominator of function $G(s)$ in Fig.~\ref{Linearly_controlled_system}.

\subsection{Root locus and pole sensitivity to parameter $K$.}

The movement of the poles of the feedback system $G_0(s)$ in the complex plane when the parameter $K$ ranges from $0$ to $\infty$ (or from $0$ to $-\infty$)
is well described by the root locus. However, the root locus
does not clearly solve the pole sensitivity problem in Definition~\ref{definition_sensitivity}, namely it does not
show the velocity $\frac{dp_j}{dK}$ of the poles $p_j$ with respect to $K$, that is how fast the poles $p_j$ move in the complex plane when the control parameter changes from $K$  to $K+dK$, where $dK$ is an infinitesimal increment of $K$. The pole sensitivity problem can be addressed using the following theorem:

\vspace{1mm}
\begin{Theo}\label{Prop_1} (Simple poles) The velocity $\frac{dp_{j}}{dK}$ in the complex plane of the poles $p_j$ of the feedback system shown in Fig.~\ref{Linearly_controlled_system}, in the case of simple poles $p_j$ and when the parameter $K$ changes from $K$ to $K+dK$, is the following:
\begin{equation}
\label{final_dp_h_over_dK}
 \frac{dp_{j}}{dK}= -\bar{k}_{j},
 \hspace{10mm}
 \mbox{ for }  j\in\{1,\,2,\,\ldots,\,n\},
\end{equation}
where $\bar{k}_{j}$ defined in \eqref{Residues} are the residues associated with the poles $p_{j}$ of the function $G_0(s)$ defined in \eqref{G0s}.
\end{Theo}
\vspace{1mm}
{\it Proof.} Using \eqref{Gs_function} and replacing $K$ with $K+dK$, the characteristic equation \eqref{equazione_caratteristica} can  be rewritten as follows:
\begin{equation}
\label{Char_Eq}
\Delta(s,K) +dK\,N(s)=0,
\end{equation}
where $\Delta(s,K)$ is the characteristic equation of the feedback system $G_0(s)$.
When $dK=0$, the roots $p_h$ of the characteristic equation \eqref{Char_Eq} are equal to the poles $p_j$  of the feedback system  in Fig.~\ref{Linearly_controlled_system} for  the given parameter $K$. When $dK \neq 0$, the new roots $\bar{p}_h$ of the characteristic equation \eqref{Char_Eq} can be expressed as $\bar{p}_h=p_{h}+dp_{h}$, where $dp_{h}$ is the infinitesimal distance of the new root $\bar{p}_h$ from the previous root $p_{h}$. Therefore, when $dK\not=0$, the  characteristic equation \eqref{Char_Eq} can be expressed as follows:
\begin{equation}
\label{Char_Eq_dK}
\Delta(s,K)+dK\,N(s)
=
\prod_{h=1}^{n}(s-p_h-dp_h)=0.
\end{equation}
The relation \eqref{Char_Eq_dK} holds $\forall s\in\mathbb{C}$.
Substituting $s=p_{j}$ in \eqref{Char_Eq_dK}, for $j\in\{1,\,2,\,\ldots,\,n\}$,
yields:
\begin{equation}
\label{Char_Eq_s_eq_p_h}
\underbrace{
\Delta(p_j,K)
}_{0} + dK N(p_j\!)
\!=\!
-dp_j\!\!\prod_{h=1,h\not=j}^{n}\!\!(p_j-p_h-dp_h),
\end{equation}
where $
\Delta(p_j,K)
=0$
because $p_j$ is a pole of the function $G_0(s)$.
Letting the infinitesimal quantities $dp_{h}$, $dp_{j}$ and $dK$ tend to zero, from \eqref{Char_Eq_s_eq_p_h}  and using the definition of residues $\bar{k}_{j}$ given in \eqref{Residues}, it follows that:
\[\frac{dp_j}{dK}
=
- \frac{N(p_j)}{\prod_{h=1,h\not=j}^{n}(p_j-p_h)}
= -\bar{k}_{j},
\]
 which is the relation \eqref{final_dp_h_over_dK} given in Throrem~\ref{Prop_1}. Relation \eqref{final_dp_h_over_dK} can be interpreted as follows: the velocities $\frac{dp_j}{dK}$ in the complex plane of the roots $p_{j}$ of the characteristic equation \eqref{equazione_caratteristica} when the parameter $K$ changes from $K$ to $K+dK$ are given by the residues $\bar{k}_{j}$ of the function $G_0(s)$ at the points $s=p_{j}$ taken with opposite signs.
 $\hfill \Box$
\vspace{1mm}
\begin{Prop}\label{Prop_2} (Multiple poles) The modulus of the velocity $\frac{dp_{j}}{dK}$
of the poles $p_j$ of the feedback system $G_0(s)$ in \eqref{G0s} in correspondence of poles $p_j$ with  multiplicity $r_j(p_{j})>1$
is infinite:
\begin{equation}
\label{final_dp_h_over_dK_hnfty}
 \left|\frac{dp_{j}}{dK}\right|= \infty
 \hspace{12mm}
 \mbox{ if }  \hspace{2mm} r_j(p_{j})>1.
\end{equation}
\end{Prop}
\vspace{1mm}
\vspace{1mm}
{\it Proof.} Relation \eqref{Char_Eq_dK} holds true for poles $p_j$ with  multiplicity $r_j(p_{j})>1$ as well. Substituting $s=p_{j}$ in \eqref{Char_Eq_dK}, where $p_{j}$ is a pole of function $G_0(s)$ with multiplicity $r_j(p_{j})>1$,  and recalling that $D(p_j\!)\!+\!KN(p_j\!)=0$ from \eqref{equazione_caratteristica}, yields:
\begin{equation}
\label{Char_Eq_s_eq_p_h_r_j}
 dK N(p_j\!)
=
(-1)^{r_j} \!\!\!\!
\prod_{h=1,p_h=p_j}^{n} \!\!\!\!dp_h
\prod_{h=1,p_h\not=p_j}^{n}\!\!\!\!(p_j-p_h-dp_h).
\end{equation}
In the vicinity of a multiple pole $p_{j}$, the  differentials $dp_{h}$ are a balanced set of vectors, with the same modulus $|dp_j|$, satisfying  the following relation:
\begin{equation}
\label{prod_p_j}
\prod_{h=1,p_h=p_j}^{n} \!\!\!\!dp_h
= (-1)^{r_j-1} |dp_j|^{r_j}.
\end{equation}
 Substituting \eqref{prod_p_j} in \eqref{Char_Eq_s_eq_p_h_r_j} and  letting the infinitesimal quantities $dp_{h}$, $dp_{j}$ and $dK$ tend to zero, yields:
\begin{equation}
\label{dp_h_over_dK_r_j}
\frac{|dp_j|^{r_j}}{dK}
=
- \frac{N(p_j)}{\prod_{h=1,p_h\not=p_j}^{n}\!(p_j-p_h)}
= -\bar{k}_{j,r_j},
\end{equation}
where $\bar{k}_{j,r_j}$ is a constant  defined  
as follows:
\begin{equation}
\label{bark_j_r_j}
\!\!\bar{k}_{j,r_j}
\!= \!\left.G_0(s)(s-p_j)^{r_j}\right|_{s=p_j}
\!=\!\frac{N(p_j)}{\prod_{h=1,p_h\not=p_j}^{n}\!(p_j-p_h)}.
\end{equation}
Dividing both sides of \eqref{dp_h_over_dK_r_j} by $|dK|^{r_j-1}$ and taking the modulus yields:
\begin{equation}
\label{dp_k_over_dK}
\left|\frac{dp_{j}}{dK}\right|
= \left(\frac{|\bar{k}_{j,r_j}|}{|dK|^{r_j-1}}\right)^\frac{1}{r_j}.
\end{equation}
Relation \eqref{bark_j_r_j} shows that, when $dK\rightarrow0$, the modulus of the velocity $\frac{dp_{j}}{dK}$ tends to infinity if the multiplicity $r_j(p_{j})$ of pole $p_j$ is grater that 1, as stated in \eqref{final_dp_h_over_dK_hnfty}. Relation \eqref{dp_k_over_dK} can be rewritten in logarithmic scales as follows:
\[\left|\frac{dp_{j}}{dK}\right|_{\log_{10}}
= \frac{|\bar{k}_{j,r_j}|_{\log_{10}}}{r_j}- \frac{(r_j-1)}{r_j}|dK|_{\log_{10}},
\]
resulting in a straight line in the logarithmic plane with a slope $t=- \frac{(r_j-1)}{r_j}$ for increasing values of $|dK|_{\log_{10}}$.
$\hfill \Box$

\vspace{1mm}

\subsubsection*{Numerical Case Study}
Reference is made to the following transfer function:
\begin{equation}\label{num_es_1}
G(s)=\frac{4\,s}{(s+4)[(s+1)^2+2^2]}.
\end{equation}
The closed-loop transfer function $G_0(s)$ is the following:
 \[
G_0(s)=\frac{4\,K\,s}{(s+4)[(s+1)^2+2^2]+4\,K\,s},
\]
whose poles when $K=0$ are $p_{1}=-4$ and $ p_{2,3}=-1\pm2\,j$. The velocity vectors  of the poles $p_1$ and $p_{2,3}$ when $K=0$ can be obtained by means of Throrem~\ref{Prop_1}
by computing the residues of function $G_0(s)$ when $K=0$:
\[
\left.\frac{dp_{1}}{dK}\right|_{K=0}\!\!\!\!\!=\!-\bar{k}_{1}\!=\!1.231,
\hspace{1.5mm}
\left.\frac{dp_{2,3}}{dK}\right|_{K=0}
\!\!\!\!\!=\!-\bar{k}_{2,3}\!=\!-0.615 \pm 0.077\,j.
\]
%
\begin{figure}
\psfrag{Real Axis}[][][0.8]{Real Axis}
\psfrag{Imag Axis}[][][0.8]{Imaginary Axis}
\psfrag{K=[0  2]}[b][b][0.8]{}
\psfrag{K01}[b][t][0.6]{$K\!=\!0$}
\psfrag{K02}[b][t][0.6]{$K\!=\!0$}
\psfrag{K03}[b][t][0.6]{$K\!=\!0$}
\psfrag{K21}[b][t][0.6]{$K\!=\!2$}
\psfrag{K22}[lb][tl][0.6]{$K\!=\!2$}
\psfrag{K23}[rb][tl][0.6]{$K\!=\!2$}
\psfrag{P01}[t][b][0.8]{$\frac{dp_1}{dK}$}
\psfrag{P02}[t][b][0.8]{$\frac{dp_2}{dK}$}
\psfrag{P03}[t][b][0.8]{$\frac{dp_3}{dK}$}
\psfrag{P21}[t][b][0.8]{$\frac{dp_1}{dK}$}
\psfrag{P22}[tr][b][0.8]{$\frac{dp_2}{dK}$}
\psfrag{P23}[tl][l][0.8]{$\frac{dp_3}{dK}$}
\centering
\includegraphics[width=0.96\columnwidth]{\eps/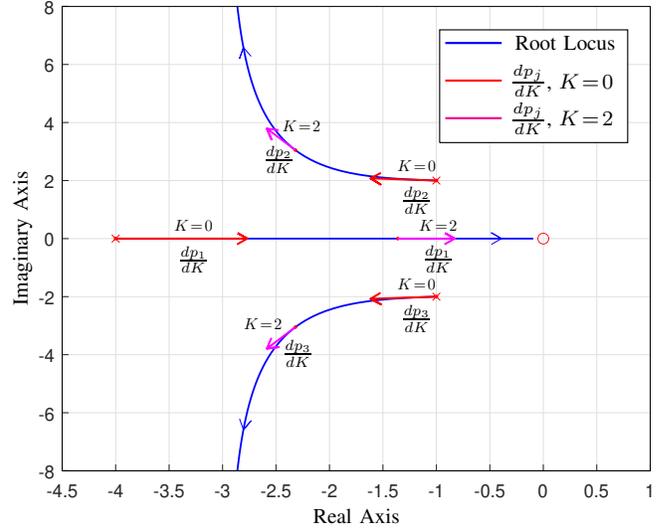}
   \setlength{\unitlength}{5.0mm}
 \psset{unit=\unitlength}
  \rput(-1.75,1.25){
  \psline[linewidth=0.42pt,fillcolor=white,fillstyle=solid](-4.5,8.9)(0.5,8.9)(0.5,11.9)(-4.5,11.9)(-4.5,8.9)
  \psline[linecolor=blue](-4.35,11.55)(-2.85,11.55)
  \rput(-1.275,11.55){ \footnotesize Root Locus}
  \psline[linecolor=red](-4.35,10.55)(-2.85,10.55)
  \rput(-1.255,10.55){\footnotesize $\frac{d p_j}{dK}$, $K\!=\!0$}
  \psline[linecolor=magenta](-4.35,9.55)(-2.85,9.55)
  \rput(-1.255,9.55){\footnotesize $\frac{d p_j}{dK}$, $K\!=\!2$}
  }
\vspace{-1mm}
\caption{Root locus and poles velocity vectors $\frac{d p_j}{dK}$ of the feedback system associated with $G(s)$ in \eqref{num_es_1}.}
\label{verifica_numerica_2_dp_dk}
\end{figure}
The poles of function $G_0(s)$ when $K=2$ are
$p_{1}=-1.362$ and $ p_{2,3}=-2.319 \pm 3.050 j$. The velocity vectors  of the poles $p_1$ and $p_{2,3}$ when $K=2$ can be obtained by means of Throrem~\ref{Prop_1}
by computing the residues of function $G_0(s)$ when $K=2$:
\[\left.\frac{dp_{1}}{dK}\right|_{K=2}\!\!\!\!\!=\!-\bar{k}_{1}\!=\!0.533,
\hspace{1.5mm}
\left.\frac{dp_{2,3}}{dK}\right|_{K=2}
\!\!\!\!\!=\!-\bar{k}_{2,3}\!=\!-0.267 \pm 0.739\,j.
\]
A graphical representation of the root locus and of $\frac{dp_{j}}{dK}$ is given in Fig.~\ref{verifica_numerica_2_dp_dk}, from which it can be seen that $\frac{dp_{j}}{dK}$ are indeed the velocity vectors of the poles $p_{j}$ as they are tangential to the root locus.

\section{Algorithm for The Root locus calculation}\label{Algorithm_for_Root_locus_calculation_sect}

The poles of the feedback system $G_0(s)$   in \eqref{G0s} are function of the parameter $K$: $p_j=p_j(K)$.
Let $\p(K)$ denote the vector of all the poles of function $G_0(s)$:
$\p(K)=\mat{cccc}{p_1(K) & p_2(K) & \ldots & p_n(K)}\tras.$
The root locus of the feedback system $G_0(s)$ is the plotting of all the components $p_j(K)$ of vector $\p(K)$ in the complex plane when parameter $K$ ranges from 0 to $\infty$. The typical approach consists in computing the components $p_j(K)$ as the roots of the characteristic equation  $\Delta(s,K)=0$ in \eqref{equazione_caratteristica}. We propose the alternative and more efficient method
described in the following algorithm.
\vspace{1mm}
\begin{Algo}\label{Prop_3_bis} (Root locus construction) Let  $\bar{K}=(K_a:\delta k:K_b)$ be the vector of the desired values of parameter $K$ constructed with precision $\delta k$.
Let $k_i\in\bar{K}$ denote the $i$-th value of vector $\bar{K}$, for $i=\{1,\;2,\;\ldots,\;N\}$. A good estimation $\tilde{p}_j(k_i)$ of the components $p_j(k_i)$ of vector $\p(k_i)$, for $k_i\in\bar{K}$, $j\in\{1,\,2,\,\ldots,\,n\}$ and  $i=\{1,\;2,\;\ldots,\;N-1\}$, can be computed by using the following difference equations:
\begin{equation}
\label{d_p_K_dK_sol_bis_new_bis}
 \tilde{p}_j(k_{i+1})   = \tilde{p}_j(k_i) -
\left.\frac{N(s)\,\delta k +\Delta(s,k_i)}{\prod_{h=1,h\not=j}^{n}(s-\tilde{p}_h(k_i))}\right|_{s=\tilde{p}_j(k_i)}
\end{equation}
where $N(s)$ is the numerator of function $G(s)$ and $\Delta(s,k_i)$ is the characteristic equation of function $G_0(s)$  in \eqref{equazione_caratteristica} for $K=k_i$. The initial conditions $\tilde{p}_j(k_{1})$ of the difference equations \eqref{d_p_K_dK_sol_bis_new_bis} are the poles   $p_j(k_{1})$  of the feedback function $G_0(s)$ when $K=k_1$:  $\tilde{p}_j(k_{1})=p_j(k_{1})$. Note:
the proposed difference equations \eqref{d_p_K_dK_sol_bis_new_bis} is to be used for values $k_i$ corresponding to simple poles $p_j(k_i)$ of the feedback function $G_0(s)$, and can be used to construct the branches of the root locus of the feedback system $G_0(s)$ in the regions outside of the branch points.
\end{Algo}
\vspace{1mm}
\begin{figure}[t!]
\psfrag{Real Axis}[][][0.8]{Real Axis}
\psfrag{Imag Axis}[][][0.8]{Imaginary Axis}
\psfrag{Root locus}[][][0.8]{}
\psfrag{K=[0  2]}[b][b][0.8]{$K=[0,\;2]$}
\psfrag{K01}[b][t][0.6]{$K\!=\!0$}
\centering
\includegraphics[width=0.8\columnwidth]{\eps/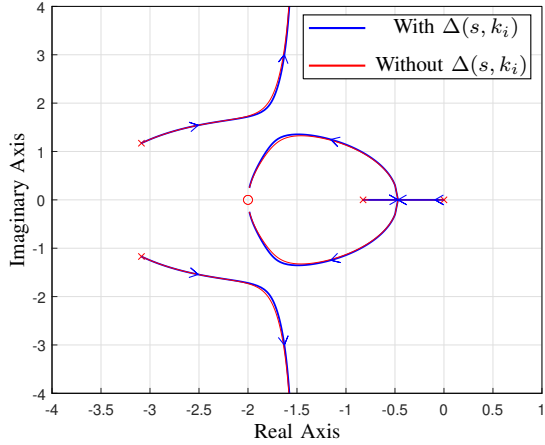}
   \setlength{\unitlength}{5.0mm}
 \psset{unit=\unitlength}
  \rput(-2.5,-0.65){
  \psline[linewidth=0.42pt,fillcolor=white,fillstyle=solid](-4.5,10.2)(1.55,10.2)(1.55,11.9)(-4.5,11.9)(-4.5,10.2)
  \psline[linecolor=blue](-4.35,11.55)(-2.85,11.55)
  \rput(-0.55,11.55){ \footnotesize With $\Delta(s,k_i)$}
  \psline[linecolor=red](-4.35,10.55)(-2.85,10.55)
  \rput(-0.55,10.55){\footnotesize Without $\Delta(s,k_i)$}
  }
  \vspace{-1mm}
\caption{Root loci of the feedback system associated with $G(s)$ in \eqref{stabil_term_comp} when $K$ ranges from $0$ to $10$ computed using Algorithm~\ref{Prop_3_bis} with and without the stabilizing term $\Delta(s,k_i)$ in \eqref{d_p_K_dK_sol_bis_new_bis}. }
\label{Root_Locus_Fun_Examples_11}
\end{figure}
{\it Proof}.
 With reference to Eq.~\eqref{Char_Eq_dK}, substituting $K$ with $k_i$,  $p_h$ with $\tilde{p}_h(k_i)$  and substituting the infinitesimal quantities $dK$ and $dp_h$ with the small quantities $\delta k$ and $\delta \tilde{p}_h(k_i)$,  yields:
\begin{equation}
\label{Char_Eq_dK_due}
\Delta(s,k_i)
+\delta k N(s)
=
\prod_{h=1}^{n}(s-\tilde{p}_h(k_i)-\delta \tilde{p}_h(k_i))=0,
\end{equation}
where $\delta \tilde{p}_h(k_i)$ is the small distance between the pole $p_h(k_{i+1})$ and the estimated pole $\tilde{p}_h(k_{i})$ due to the small variation  $\delta k$ of parameter $K$:
\begin{equation}
\label{delta_tilde_p}
\delta \tilde{p}_h(k_i)= p_h(k_{i+1})-\tilde{p}_h(k_i).
\end{equation}
Relation \eqref{Char_Eq_dK_due} holds $\forall s\in\mathbb{C}$.
Substituting $s=\bar{s}=\tilde{p}_{j}(k_i)$ in \eqref{Char_Eq_dK_due} yields:
\[\Delta(\bar{s},k_i)
+\delta k N(\bar{s})
\!=\!
-\delta \tilde{p}_j(k_i)\!\!\!
\prod_{h=1,i\not=j}^{n}\!\!\!(\bar{s}-\tilde{p}_h(k_i)-\delta \tilde{p}_h(k_i))\!=\!0.
\]
From
the latter and from \eqref{delta_tilde_p}, when $h=j$, it follows that:
\begin{equation}
\label{dp_h_over_dK_due}
p_j(k_{i+1})=\tilde{p}_j(k_i)
- \frac{
N(\bar{s})\delta k +\Delta(\bar{s},k_i)}{\prod_{h=1,h\not=j}^{n}(\bar{s}-\tilde{p}_h(k_i)-\delta \tilde{p}_h(k_i))}.
\end{equation}
A good estimation $\tilde{p}_j(k_{i+1})$ of the pole $p_j(k_{i+1})$  when $K=k_{i+1}$  can be obtained from \eqref{dp_h_over_dK_due} by neglecting, for $h\not=j$, the small quantities $\delta \tilde{p}_h(k_i)$ with respect to the values $\bar{s}-\tilde{p}_h(k_i)$:
\[\tilde{p}_j(k_{i+1})=\tilde{p}_j(k_i)
- \frac{
N(\bar{s})\delta k +\Delta(\bar{s},k_i)}{\prod_{h=1,h\not=j}^{n}(\bar{s}-\tilde{p}_h(k_i))},
\]
 which coincides with the difference equation \eqref{d_p_K_dK_sol_bis_new_bis}. $\hfill \Box$

 A dedicated MATLAB function has been created for the implementation of Algorithm~\ref{Prop_3_bis}, which is publicly available at the repository reported in~\cite{rloc_repository}.

\vspace{1mm}

Note: an important stabilizing term of Eq.~\eqref{d_p_K_dK_sol_bis_new_bis} is the term $\Delta(s,k_i)$ with $s=\bar{s}=\tilde{p}_{j}(k_i)$. In fact, if the estimated pole $\tilde{p}_{j}(k_i)$ belongs to the root locus of 
$G_0(s)$, i.e. if $\tilde{p}_{j}(k_i)=p_{j}(k_i)$, then $\Delta(\tilde{p}_{j}(k_i),k_i)=0$ does not influence Eq.~\eqref{d_p_K_dK_sol_bis_new_bis}.  On the contrary,  if the estimated pole $\tilde{p}_{j}(k_i)$ does not belong to the root locus of $G_0(s)$, the term $\Delta(s,k_i)$ provides a correction action to Eq.~\eqref{d_p_K_dK_sol_bis_new_bis} in the correct direction, which is proportional to the distance of the estimated pole $\tilde{p}_{j}(k_i)$ from the actual root locus. Therefore, the term  $\Delta(s,k_i)$ always keeps the next estimated pole $\tilde{p}_{j}(k_{i+1})$ in the vicinity of the actual root locus.

To further verify the effectiveness of the stabilizing term $\Delta(s,k_i)$, a comparative simulation has been performed. The root locus of the feedback system $G_0(s)$ in Fig.~\ref{Linearly_controlled_system} with the following open-loop transfer function
\begin{equation}\label{stabil_term_comp}
G(s)= \frac{10(s+2)^2}{s^2(s+2)^2+3 s (s+1) (s+3)}
\end{equation}
has been constructed using the proposed Algorithm~\ref{Prop_3_bis} with and without the stabilizing term $\Delta(s,k_i)$ when $K$ ranges from $0$ to $10$ and using $\delta k=0.01$, and the results are shown in Fig.~\ref{Root_Locus_Fun_Examples_11}. The MATLAB {\tt rlocus} function has been employed as ground truth. The average error $Err_{maen}$ and the maximum error $Err_{max}$ with respect to the ground truth when using the stabilizing term $\Delta(s,k_i)$ are $Err_{maen}=6.13 \cdot 10^{-5}$ and  $Err_{max}=3.5\cdot 10^{-3}$ while,
%
%
without using the stabilizing term $\Delta(s,k_i)$, they are $Err_{maen}=5.11 \cdot 10^{-3}$ and  $Err_{max}=5.07\cdot 10^{-2}$, which are respectively two orders of magnitude and one order of magnitude larger than the previous ones, thus showing the effectiveness of the stabilizing term $\Delta(s,k_i)$.
 \begin{figure}[t!]
  \centering
 \psfrag{Avg Exec Time}[][t][1]{Average Execution Time [s]}
\psfrag{G(s)}[][][1]{$G(s)$}
  \includegraphics[clip,width=\columnwidth]{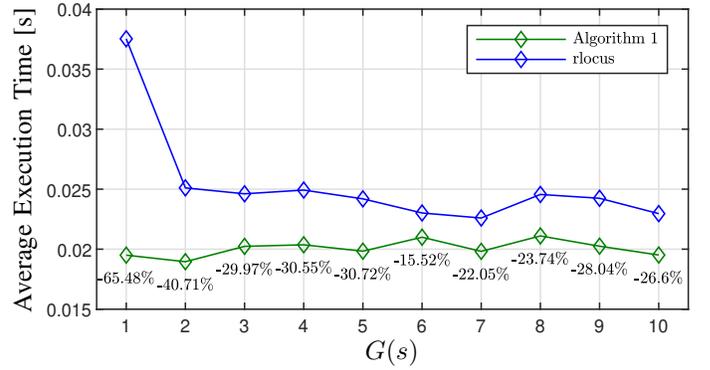}
  \vspace{-6mm}
 \caption{Comparison between the average execution times of Algorithm~\ref{Prop_3_bis} and {\tt rlocus} for the construction of the root locus.}
 \label{Root_Locus_Fun_Examples_MIO_2}
\end{figure}

\subsection{Comparison with the MATLAB root locus function}
The Algorithm~\ref{Prop_3_bis} for the construction of the root locus has been tested against the MATLAB {\tt rlocus} function to evaluate their performances. The two algorithms have been tested over ten different transfer functions, reported in the following:
\[
\begin{array}{cc}
G_1(s)=\frac{s+3}{s(s+2)}, &
G_2(s)=\frac{4\,s}{(s+4)((s+1)^2+2^2)}, \\[1.5mm]
G_3(s)=\frac{10(s-1)}{s(s+1)(s^2+8s+25)},&
G_4(s)=\frac{10(s + 0.2)}{(s - 1)(s - 2)(s + 10)},\\[1.5mm]
G_5(s)=\frac{0.8}{s((s+2)^2+1)},&
G_6(s)=\frac{12}{s(s+1)(s+2)(s+3)},\\[1.5mm]
G_7(s)=\frac{12}{s((s+2)^2+1)(s+4)},&
G_8(s)=\frac{10(s+2)^2}{s^2(s+4)^2},\\[1.5mm]
G_9(s)=\frac{30(s^2+4s+25)}{s^2(s + 2)(s + 6)},&
G_{10}(s)=\frac{10(s+2)^2}{s^2(s+2)^2+3s(s+1)(s+3)}.
\end{array}
\]
For each transfer function, the root locus has been computed ten times using Algorithm~\ref{Prop_3_bis} and the MATLAB {\tt rlocus} function in the same nominal conditions, and the average execution times have been reported in Fig.~\ref{Root_Locus_Fun_Examples_MIO_2}. The obtained results clearly show that the proposed Algorithm~\ref{Prop_3_bis} is more efficient than the MATLAB {\tt rlocus} function, given that the average execution time of Algorithm~\ref{Prop_3_bis} is from $15\%$ to $65\%$ lower than that of the MATLAB {\tt rlocus} function.

\section{Pole sensitivity of a dynamical system to its own parameters}\label{Poles_sensitivity_of_a_dynamic_system_to_its_own_parameters_sect}
Let $\cS(\h)$ be a linear dynamical physical system, and let $\h$ denote the vector of all its parameters $h_i$, for $i\in\{1,\,2,\,\ldots,\,p\}$:
$\h=\mat{cccc}{h_1 & h_1 & \ldots & h_p}\tras.$
 The parameters $h_i$ of a physical system $\cS(\h)$ can be distinguished in the following three different types~\cite{POG_1,POG_2}:
\\
1) The {\it dynamic parameters}
are those that can only {\it store} energy (example: capacitor);
\\
2) The {\it static parameters}
are those that can only {\it dissipate} energy (example: resistance);
\\
3) The {\it connection parameters}
are those that can only {\it convert} energy (example: the torque constant of a DC motor).
\\[1mm]
The pole sensitivity of the dynamical  system $\cS(\h)$  to its own parameters can be performed if the characteristic polynomial $\Delta(s,\h)$ of the system $\cS(\h)$ is known, which can be obtained in many different ways. The two easiest ways to obtain the  polynomial $\Delta(s,\h)$ are the following:
\\[1mm]
1) $\Delta(s,\h)$ is the denominator of any transfer function $G_{ij}(s,\h)$ calculated between two generic input-output points $(i,\,j)$ of the considered system $\cS(\h)$:
\[
G_{ij}(s,\h)
= \frac{Y_j(s)}{X_i(s)}
= \frac{N_{ij}(s,\h)}{\Delta(s,\h)},
\]
where $X_i(s)$ and  $Y_j(s)$ are the Laplace transform of the $i$-th input signal $x_i(t)$ and of the $j$-th output signal $y_j(t)$ of the considered system, and the polynomial  $N_{ij}(s,\h)$ is the numerator of the transfer function $G_{ij}(s,\h)$.
\\[1mm]
2) If the block scheme of the considered system $\cS(\h)$ is available, then the characteristic polynomial $\Delta(s,\h)$ is the numerator of the Mason determinant $M(s)$ of the block scheme:
\[
M(s,\h)
= \frac{\Delta(s,\h)}{D_M(s,\h)},
\]
 where the polynomial $D_M(s,\h)$ is the denominator of the
 transfer function $M(s,\h)$.
\vspace{1mm}

\begin{Prop}\label{Prop_4} (Characteristic polynomial) All the physical parameters $h_i$ of system
$\cS(\h)$ appear within the characteristic polynomial  $\Delta(s,\h)$ in the following way:
a) all the {\it dynamic} and {\it static parameters} $h_i$ of system $\cS$ appear {\it linearly} within the polynomial  $\Delta(s,\h)$;
b) the  {\it squared forms $h_i^2$} of all the {\it connection parameters} of system $\cS$ appear {\it linearly} within the polynomial  $\Delta(s,\h)$.
\end{Prop}
\vspace{1mm}

{\it Proof.} This Property can be proven by referring to the following basic properties of the main power-oriented modeling techniques known in the literature, namely Power-Oriented Graphs (POG)~\cite{POG_1,POG_2}, Bond Graphs (BG)~\cite{BG_1} and Energetic Macroscopic Representation (EMR)~\cite{EMR_1}:  1) Any physical system can always be   obtained by connecting in series or in parallel the physical elements composing it; 2) When a new physical element is connected to a linear system, a new feedback loop appears within the mathematical model of the system; 3) If the new connected  element is of {\it dynamic} or {\it static} type, then the corresponding parameter $\bar{h}_i$ appears linearly both in the new feedback loop and in the characteristic polynomial $\Delta(s)$ of the  new system; 4) On the contrary, if the new inserted element is of {\it connection}  type,
then the corresponding parameter $\bar{h}_i$ appears twice in the mathematical model of the considered system, because an energy conversion takes place.
This
induces the square $\bar{h}_i^2$ of the physical parameter $\bar{h}_i$ to appear linearly within the characteristic polynomial $\Delta(s)$ of the considered system.
$\hfill \Box$
 \\[1mm]
 When the $i$-th parameter $h_i$ of vector $\h$ changes from $h_i$  to $h_i+dh_i$, for $i\in\{1,\,2,\,\ldots,\,p\}$, the characteristic equation $\Delta(s,h_i)=0$ of system $\cS(\h)$ can always be written using the following truncated Taylor expansion:
\begin{equation}
\label{Delta_s_Form}
\Delta(s,h_i+dh_i)\simeq\Delta(s,h_i)+ dh_i \bar{N}_i(s,h_i)=0,
\end{equation}
where polynomial $\bar{N}_i(s,h_i)$ can be computed as follows:
\begin{equation}
\label{Delta_s_Form_bisss}
\bar{N}_i(s,h_i)= \frac{\partial \Delta(s,h_i)}{\partial h_i}.
\end{equation}
\vspace{1mm}

\begin{Theo}\label{Prop_5} (Pole sensitivity to system's parameters) The velocity $\frac{dp_{j}}{dh_i}$ in the complex plane of the poles $p_j$ of the system $\cS(\h)$,
when the parameter $h_i$ changes from $h_i$  to $h_i+dh_i$, is the following:
\begin{equation}
\label{final_dp_h_over_dK_bis}
 \frac{dp_{j}}{dh_i}= -\tilde{k}_{ji},
 \hspace{6mm}
 \mbox{ for }
 j\in\{1,\,2,\,\ldots,\,n\},
\end{equation}
where $\tilde{k}_{ji}$ are the residues computed as in \eqref{Residues} of the following transfer functions $G_i(s)$:
\begin{equation}
\label{G_i_s}
G_i(s,h_i)
= \frac{\bar{N}_i(s,h_i)}{\Delta(s,h_i)},
\end{equation}
where $\bar{N}_i(s,h_i)$ is the polynomial defined in \eqref{Delta_s_Form_bisss}.
\end{Theo}
\vspace{1mm}
 {\it Proof.}
The relations \eqref{final_dp_h_over_dK_bis} and \eqref{G_i_s} can be derived by applying Theorem~\ref{Prop_1} to the characteristic equation \eqref{Delta_s_Form}.
$\hfill \Box$
\\[1mm]

\subsubsection*{Physical System Case Study (DC electric motor)}
\begin{figure}
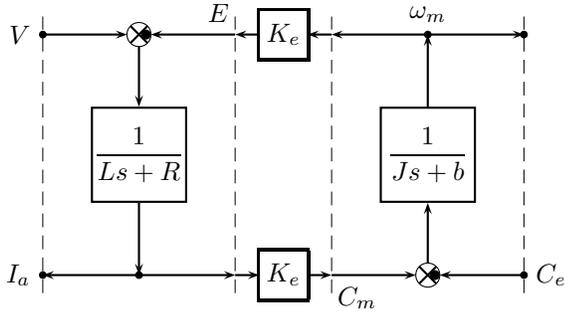

\centering
 \SpecialCoor
 \schemaPOG{(0,-1)(20,12)}{3.2mm}{
  \thicklines
  \bloin{V}{I_{a}}
  \blogiu{\ds \frac{1}{Ls+R}}{}{\dst}
  \blokao{K_{e}}{K_{e}}{E}{C_{m}}
  \blosu{\ds \frac{1}{Js+b}}{\omega_{m}}{\dst}
  \bloout{}{C_{e}}
}
\caption{Power-Oriented Graphs block scheme of a DC electric motor.}
\label{DC_electric_motor}
\end{figure}

\begin{figure}[t!]
\centering
\psfrag{dR1}[tr][tl][0.9]{$\frac{dp_1}{dR}$}
\psfrag{dJ1}[l][t][0.9]{$\frac{dp_1}{dJ}$}
\psfrag{dL1}[][][0.9]{$\;\frac{dp_1}{dL}$}
\psfrag{dKe1}[][][0.9]{$\;\;\;\frac{dp_1}{dK_e}$}
\psfrag{db1}[br][b][0.9]{$\frac{dp_1}{db}\;$}
\psfrag{dR2}[br][bl][0.9]{$\frac{dp_2}{dR}$}
\psfrag{dJ2}[l][b][0.9]{$\frac{dp_2}{dJ}$}
\psfrag{dL2}[][][0.9]{$\;\frac{dp_2}{dL}$}
\psfrag{dKe2}[][][0.9]{$\;\;\;\frac{dp_2}{dK_e}$}
\psfrag{db2}[tr][t][0.9]{$\frac{dp_2}{db}\;$}
\includegraphics[width=0.9\columnwidth]{\eps/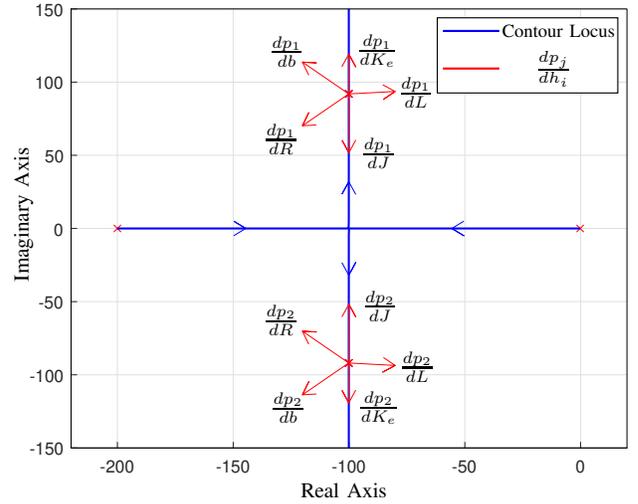}
   \setlength{\unitlength}{5.0mm}
 \psset{unit=\unitlength}
  \rput(-1.2,0.25){
  \psline[linewidth=0.42pt,fillcolor=white,fillstyle=solid](-4.5,9.9)(0.3,9.9)(0.3,11.9)(-4.5,11.9)(-4.5,9.9)
  \psline[linecolor=blue](-4.35,11.55)(-2.85,11.55)
  \rput(-1.4,11.55){ \scriptsize Contour Locus}
  \psline[linecolor=red](-4.35,10.55)(-2.85,10.55)
  \rput(-1.4,10.55){\footnotesize $\frac{d p_j}{dh_i}$}
\rput(-6.98,-0.68){\footnotesize Real Axis}
\rput{90}(-15.5,6.68){\footnotesize Imaginary Axis}
  }
\vspace{1.68mm}
\caption{Contour locus as a function of parameter $K_e$ and poles velocity vectors $\frac{d p_j}{dh_i}$ of the DC motor in Fig.~\ref{DC_electric_motor} due to the variation of the system parameters $h_i\in\h$.}
\label{var_par_motore}
\end{figure}

%
The POG block scheme~\cite{POG_1,POG_2} of a DC electric motor is shown in Fig.~\ref{DC_electric_motor}.
 The  vector $\h$ of the system's parameters is
$\h=\mat{ccccc}{L & R & J & b & K_e}\tras$,
 where the inductance $L$ and inertia $J$ are dynamic parameters, the resistance $R$ and the linear friction coefficient $b$ are static parameters, and the motor  torque constant $K_e$ is a connection parameter, as described in Sec.~\ref{Poles_sensitivity_of_a_dynamic_system_to_its_own_parameters_sect}.
The Mason determinant of the POG scheme in Fig.~\ref{DC_electric_motor} is:
\[
M(s)
= 1+\frac{K_e^2}{(Ls+R)(Js+b)}
= \frac{\Delta(s,\h)}{D(s)}.
\]
The characteristic equation of the system is the following:
\begin{equation}
\label{Determinante_grafo}
\Delta(s)=(L\,s\!+\!R)(J\,s\!+\!b)\!+\!K_{e}^{2}=0,
\end{equation}
which satisfies Property~\ref{Prop_4}.
By applying Theorem~\ref{Prop_5} to the characteristic equation \eqref{Determinante_grafo}, the velocity $\frac{dp_{j}}{dh_i}$ in the complex plane of the poles $p_j$ of the considered DC motor, when parameters $h_i$ change from $h_i$  to $h_i+dh_i$, is completely defined
by the residues $\tilde{k}_{ji}$ of the following transfer functions $G_i(s,h_i)$:
\[
\begin{array}{c}
 G_1(s,L)
= \frac{1}{\Delta(s)}\frac{\partial \Delta(s)}{\partial L}
=  \frac{s(J\,s+b)}{(L\,s\!+\!R)(J\,s\!+\!b)\!+\!K_{e}^{2}},
\\[3mm]
 G_2(s,R)
= \frac{1}{\Delta(s)}\frac{\partial \Delta(s)}{\partial R}
=  \frac{(J\,s+b)}{(L\,s\!+\!R)(J\,s\!+\!b)\!+\!K_{e}^{2}},
\\[3mm]
 G_3(s,J)
= \frac{1}{\Delta(s)}\frac{\partial \Delta(s)}{\partial J}
=  \frac{s(L\,s+R)}{(L\,s\!+\!R)(J\,s\!+\!b)\!+\!K_{e}^{2}},
\\[3mm]
 G_4(s,b)
= \frac{1}{\Delta(s)}\frac{\partial \Delta(s)}{\partial b}
=  \frac{(L\,s+R)}{(L\,s\!+\!R)(J\,s\!+\!b)\!+\!K_{e}^{2}},
\\[3mm]
G_5(s,K_e)
= \frac{1}{\Delta(s)}\frac{\partial \Delta(s)}{\partial K_e}
=  \frac{2\,K_{e}}{(L\,s\!+\!R)(J\,s\!+\!b)\!+\!K_{e}^{2}}.
\end{array}
\]
A graphical representation of the poles velocity vectors $\frac{dp_{j}}{dh_i}$
of the DC motor as a function of the variations $dh_i$ of the  system parameters $h_i\in\h$ is shown in Fig.~\ref{var_par_motore}, when the system parameters are: $L=0.005$, $R=1$, $J=0.010$, $b=0.002$, and $K_e=0.96$.
The modules of the poles velocity vectors $\frac{dp_{j}}{dh_i}$ in Fig.~\ref{var_par_motore} have been rescaled by the following coefficients $[0.2\; 40\; 40\; 0.4\; 5000]$ in order to make them visible in the figure.

\section{Conclusion}\label{Conclusion_sect}

This letter has provided a new interpretation of residues in the partial fraction decomposition, which has been employed as a basis for different applications. The first one is the study of how the speed of the poles of a feedback system change in the complex plane as a function of the control parameter, as well as the study of how the speed of the poles change when the system is subject to parameters variations. The second application is the proposal of a new algorithm for the construction of the root locus, which is proven to be more efficient in terms of execution time than the dedicated MATLAB function, while providing the same output results.

\end{document}